\documentclass{elsart}
\usepackage{graphicx,harvard,amssymb}
\journal{New Astronomy}

\def\mincir{\raise -2.truept\hbox{\rlap{\hbox{$\sim$}}\raise5.truept
\hbox{$<$}\ }}
\def\magcir{\raise -4.truept\hbox{\rlap{\hbox{$\sim$}}\raise5.truept
\hbox{$>$}\ }}

\begin{document}
\begin{frontmatter}

\title{Limber equation for luminosity dependent correlations}  

\author[dip,infn,cilea]{A. Gardini\thanksref{ag}}
\author[bic,infn]{S.A. Bonometto\thanksref{sb}}
\author[dip]{A. Macci\`{o}\thanksref{am}}

\thanks[ag]{E-mail: gardini@mi.infn.it}
\thanks[sb]{E-mail: bonometto@mi.infn.it}
\thanks[am]{E-mail: andrea@axrialto.uni.mi.astro.it}

\address[dip]{Dipartimento di Fisica, Universit\`{a} di Milano,
Via Celoria 16, I-20133 Milano, ITALY}

\address[bic]{Dipartimento di Fisica G. Occhialini, Universit\`{a} 
di Milano--Bicocca, ITALY}

\address[infn]{INFN sezione di Milano, Via Celoria 16, I-20133 Milano, ITALY}

\address[cilea]{CILEA -- Segrate (Milano), ITALY}

\begin{abstract}
The passage from angular to spatial correlations, in the case of spatial
clustering length depending on the average distance between nearby objects
is studied. We show that, in a number of cases, the scaling law
of angular correlation amplitudes, which holds for constant spatial
clustering length, is still true also for a luminosity dependent
spatial correlation. If the Limber equation is then naively
used to obtain `the' spatial clustering length from the angular
function amplitude, a quantity close to the average object separation is
obtained. The case of cluster clustering is explicitly considered.
\end{abstract}


\end{frontmatter}

\section{Introduction}

Initial studies on clustering properties over large scales were performed
using 2--dimensional samples (Totsuji \& Kihara 1969, Peebles 1980; 
see also Sharp, Bonometto \&  Lucchin 1984). Spatial properties were
then deduced from angular ones
using the Limber equation (Limber 1953). However, even in the present 
epoch of large redshift samples, the technique used to pass from angular 
to spatial statistical properties is far from becoming obsolete.
For instance, cluster samples are available and in preparation, selected
with a fixed apparent magnitude limit. Even though the typical
redshift of each cluster is given, the different intrinsical luminosity
limit at different distances forces to cut off some information, to extract
volume and intrinsical magnitude limited subsamples, where the spatial
2--point function can be directly estimated. Working out the angular
function and using the Limber equation to extract spatial properties, might
then seem a more convenient alternative approach. This approach was followed,
e.g., by Dalton et al (1997), when discussing the clustering of clusters in
catalogues extracted from the APM Galaxy Survey. In the final section
we shall comment on the results of their analysis. The point is that this
approach meets a different obstacle, as the cluster correlation length $r_c$
may depend on cluster luminosities $L$ (see Bahcall \& West 1988 for
a review of Abell cluster clustering data; see Nichol et al 1992, Dalton et
al 1994, Croft et al 1997, for APM cluster clustering data; the whole set of
data has been recently reanalysed by Lee \& Park (1999), with the inclusion
of X--ray catalog data).

Owing also to the different definition of clusters in different
observational samples or simulations, it is usual to work out from
the cumulative cluster luminosity functions $n(>L)$ the average
distance between nearby objects $D_L \equiv n^{-1/3} (>L)$
and to seek the dependence of $r_c$ on $D_L$. For 30$\, h^{-1}$Mpc$
\mincir$$ D_L$$ \mincir 60\, h^{-1}$Mpc ($h$ is the Hubble constant in
units of 100 km/s/Mpc), the Bahcall \& West (1988) conjecture that
$$
r_c \simeq 0.4 D_L
\eqno (1.1)
$$
approaches most data, although, clearly, it cannot be considered an
observational output. However, over greater scales, there is no agreement
among results obtained from
different samples and it is not clear how much this may depend on the
different cluster definitions. For Abell clusters, however, eq. (1.1)
approaches data up to $\sim 90\, h^{-1}$Mpc. In this work, we shall
assume the validity of eq.~(1.1), as an example of gross discrepancy
from assuming that $r_c$ is $L$--independent. 

In the presence of such kind of $L$--dependence, two questions arise:
(i) As a byproduct of the Limber equation, a {\sl scaling law} holds,
among the coefficients of the angular functions for samples with
different limiting magnitudes. Is such scaling law still valid?
We shall show that the same scaling law holds in the case (1.1),
as in the constant $r_c$ case.
(ii) If the Limber equation is then formally used to work out the
spatial clustering length, which value of $r_c$ is obtained?
We shall show that a naive use of the standard Limber equation
leads to working out an {\sl apparent} correlation length, close
to the average distance between nearby clusters for the whole sample.

As is obvious, our treatment applies to any kind of objects,
not to clusters only. Furthermore, the law (1.1) can be relaxed
in various ways, obtaining related results, provided that the
luminosity dependence of the clustering length ($r_c$) can be expressed
through a function of the mutual distance of the objects belonging to
an intrinsical luminosity limited set ($D_L$). Such  generalizations
will not be debated here. Our treatment also neglects departures from
euclidean geometry and $\kappa$--correction. Their inclusion does not change
the conclusions of the paper, although slightly complicating final
expressions; of course, for several detailed applications, they are
to be included, but, at this stage, they are unessential to make our points.

In order to show our points, we shall first need to reobtain
the Limber equation in a way slightly different from textbooks
(see, e.g., Peebles 1980a); this is the content of section 2 and 3.
Using algorithms and expressions deduced in these sections, in section
4 we show the points made in the Introduction.
Section 5 contains a few final remarks.

\section{Sample definition}
  
The Limber equation was introduced to work out the spatial 2--point
correlation function from angular data assuming that:
(i) the 2-point function is luminosity independent; (ii) there  
exists a universal luminosity function for the class of objects considered.
  
If the objects are galaxies, the luminosity  
function can be given the Schechter form (Schechter 1976):  
$$  
\phi (L) = {\phi^* \over L_*} \left(L \over L_*\right)^{-\alpha}  
           \exp \left(-{L \over L_*} \right)  
\eqno (2.1)  
$$  
Here $\alpha (\sim 1.1$--$1.2)$ is a phaenomenological parameter; $L_*$ is
a typical galaxy luminosity, close to the top luminosity of the sample. 
The number density of galaxies with luminosity exceeding $L$ reads  
$$  
n(>L) = \int_L^\infty dL\, \phi(L) = \phi_* U(L/L_*)
\eqno (2.2) 
$$ 
with
$$
U(\lambda) = \int_\lambda^\infty du\, u^{-\alpha} \exp(u)
\eqno (2.3)
$$
and, henceforth, $\phi(L) = -dn(>L)/dL$.

If we deal with clusters or other kinds of objects, the expression (2.1)
may no longer be a fair approximation. In the sequel, however, we shall
only need that $n(>L)$ can be expressed as in eq.~(2.2), even though
$U(\lambda)$ has not the expression (2.3). In eq.~(2.2), $\phi_*$ should
be expressed in Mpc$^{-3}h^3$ and approach the total number density for
the class of objects considered and, of course, there must also exist
a typical luminosity $L_*$, even though the distribution around it is 
not given by eq.~(2.1). As previously stated, let us then define 
$$ 
D_L \equiv n^{-1/3} (>L) ~;~~ n(>L) \equiv D_L^{-3} ~, 
\eqno (2.4) 
$$  
clearly $dn/dD = -3/D^4$.  
 
The apparent luminosity of a source of intrinsical luminosity $L$, at a  
distance $r$ from the observer, is $l = L/4\pi r^2$ and the {\sl depth} $d_*$ 
of a {\sl sample} of objects, with apparent luminosity $l > l_m$, is 
defined so that 
$$ 
l_m = L_*/4\pi d_*^2 ~. 
\eqno (2.5) 
$$ 
Hence, requiring that an object at distance $r$ has $l>l_m$, means that its 
intrinsical luminosity 
$$ 
L > L_* (r/d_*)^2 ~. 
\eqno (2.6) 
$$ 
For instance, the angular density of objects, in a sample of depth $d_*$, 
reads 
$$ 
n_\Omega (d_*) = \int_0^\infty dr\, r^2 \int_{L_*(r/d_*)^2}^\infty dL\, \phi(L) 
               = \int_0^\infty dr\, r^2 D^{-3}_{L_*(r/d_*)^2} 
               = d_*^3 G ~, 
\eqno (2.7) 
$$ 
where 
$$ 
G =  \int_0^\infty dq\, q^2 D^{-3}_{L_*q^2} 
  = {1 \over 2 L_*^{3/2}} \int_0^\infty dL \sqrt{L} D^{-3}_{L} 
\eqno (2.8) 
$$ 
is a universal constant, provided that $\phi (L)$ is a universal distribution,
quite independently from the explicit form of $U(\lambda)$. 
According to eq.~(2.7), $n_\Omega \propto d_*^3$, a well known 
volume effect.

\section{The case of luminosity independent correlations }
 
All above relations hold provided that the assumption (ii) is true. 
For a sample of objects of depth $d_*$, the angular 2--point function 
$$ 
w(\theta) = n_\Omega^{-2} 
                  \int_0^\infty dr_1 r_1^2  
                  \int_{L_*(r_1/d_*)^2}^\infty dL_1 \phi(L_1) 
                  \int_0^\infty dr_2 r_2^2 
                  \int_{L_*(r_2/d_*)^2}^\infty dL_2 \phi(L_2) 
                  \, \zeta(r_{12},L_1,L_2) 
\eqno (3.1) 
$$ 
is obtainable from the spatial 2--point functions $\zeta(r_{12},L_1,L_2)$ 
and from the luminosity function $\phi(L)$. 
 
Let us first review the standard case, when also the assumption (i) holds, 
$i.e.$ $\zeta = (r_0/r)^\gamma$ with constant (luminosity independent) 
$\gamma$ and $r_0$ (correlation length). In this case, changing integration 
variables, eq.~(3.1) immediately yields: 
$$ 
w(\theta) = n_\Omega^{-2} r_0^\gamma 
                  \int_0^\infty dr_1 r_1^2  
                  \int_{D_{L_*(r_1/d_*)^2}}^\infty dD_1 {3 \over D_1^4} 
                  \int_0^\infty dr_2 r_2^2 
                  \int_{D_{L_*(r_2/d_*)^2}}^\infty dD_2 {3 \over D_2^4} 
                  r_{12}^{-\gamma} 
\eqno (3.2) 
$$ 
and, therefore, 
$$ 
w(\theta) = n_\Omega^{-2} r_0^\gamma 
            \int_0^\infty dr_1 r_1^2 D^{-3}_{L_*(r_1/d_*)^2} 
            \int_0^\infty dr_2 r_2^2 D^{-3}_{L_*(r_2/d_*)^2} r_{12}^{-\gamma} 
\eqno (3.3) 
$$ 
Herefrom, using eq.~(2.7) and performing the changes of variables: 
$$ 
2r = r_1 + r_2 ~,~~ q = r/d_* ~,~~ ud_*\theta = r_1 - r_2 
\eqno (3.4) 
$$ 
in order that $r_{12} \simeq ud_*\theta (q^2 + u^2)^{1/2}$, we work out 
$$ 
w(\theta) = A_\gamma \left( r_0 \over d_* \right)^\gamma \theta^{1-\gamma} 
\eqno (3.5) 
$$ 
with 
$$ 
A_\gamma = c_\gamma { \int_0^\infty dq\, q^{5-\gamma}  D^{-6}_{L_*q^2} \over 
                    [ \int_0^\infty dq\, q^2 D^{-3}_{L_*q^2} ]^2 } 
         = 2\, c_\gamma \, L_*^{\gamma/2}  
                    { \int_0^\infty dL\, L^{2-\gamma/2} D^{-6}_L  \over  
                    [ \int_0^\infty dL\, \sqrt{L} D^{-3}_L ]^2 } 
\eqno (3.6) 
$$ 
where  
$$ 
c_\gamma = \int_{-\infty}^{+\infty} du\, (1+u^2)^{-\gamma/2}  
\eqno (3.7) 
$$ 
is a purely numerical constant.  
 
In above relations, the distances $D_L$ were used, instead of the number 
densities $n(>L) = D_L^{-3}$, which might even seem more convenient, 
in order to facilitate the forthcoming passage to the case of luminosity  
dependent correlations. 
 
Eq.~(3.5) puts in evidence the {\sl scaling 
properties} of the angular 2--point function. The only dependence 
on the depth of the sample ($d_*$) is factorized, as $A_\gamma$ is a 
universal expression (still, provided that $\phi (L)$ is a 
universal distribution). It is often stated that such scaling is 
arises from the fact that $r_0$ does not depend on luminosities. 
Here we shall verify that such scaling is a more general property.

\section{The case of luminosity--dependent correlations }
 
In the case of galaxy clusters, we shall now assume that
the Bahcall and West conjecture holds:
$$ 
\xi(r_{12},>L) = a\, (D_L/r_{12})^\gamma ~. 
\eqno (4.1) 
$$ 
Here the numerical constant $a \simeq 0.4$ and $D_L^{-3} = n(>L)$ is 
the number density of clusters with luminosity $>L$ (see eq.~2.5).
Eq.~(4.1) is an {\sl integral} correlation law, as it concerns
all objects with luminosities above $L$. In principle, it is possible to
define also a {\sl differential} correlation law, holding if objects
of intrinsical luminosities $L_1$ and $L_2$ only are considered.
Such differential law is not to be measured, but we want
to show that,  the integral expression (4.1) follows from assuming that
the differential law
$$ 
\zeta(r_{12},L_1,L_2)  = \zeta_* (\sqrt{D_1 D_2}/r_{12})^\gamma 
\eqno (4.2) 
$$ 
holds, provided that $a = \zeta_*(1-\gamma/6)^{-2}$. In fact 
$$ 
\xi(r_{12},>L) = n^{-2} (>L)\, \zeta_* 
                 \int_L^\infty dL_1 \phi(L_1) D_{L_1}^{\gamma/2}  
                 \int_L^\infty dL_2 \phi(L_2) D_{L_2}^{\gamma/2}  
                 r_{12}^{-\gamma} 
\eqno (4.3) 
$$ 
and, changing variables as we did to pass from eqs.~(3.1) to (3.2),
$$ 
\xi(r_{12},>L) = { \zeta_* D_L^6 \over r_{12}^\gamma } 
           \left( \int_{D_L}^\infty dD\, {3 \over D^{4-\gamma/2}} \right)^2 
         = \zeta_* (1-\gamma/6)^{-2} \left(D_L \over r_{12} \right)^\gamma 
\eqno (4.4) 
$$ 
according to eq.~(4.1). 
 
Let us now use the expression (4.2) to work out the angular function. 
In this case the correlation length has a precise dependence on 
luminosities. In spite of that, we shall find that eq.~(3.5) is 
still formally true and, in particular $w(\theta)$ has the same 
{\sl scaling properties} as in the case of luminosity independent $r_0$. 
 
In fact, if we replace the expression (4.2) in eq.~(3.1) and perform the 
operations leading there to eqs.~(3.2) and (3.3), we find 
$$ 
w(\theta) = n_\Omega^{-2} \zeta_* 
            \int_0^\infty dr_1 r_1^2  
            \int_{D_{L_*(r_1/d_*)^2}}^\infty dD_1 {3 \over D_1^{4-\gamma/2}} 
            \int_0^\infty dr_2 r_2^2 
            \int_{D_{L_*(r_2/d_*)^2}}^\infty dD_2 {3 \over D_2^{4-\gamma/2}} 
            r_{12}^{-\gamma} 
$$$$          = n_\Omega^{-2} \zeta_* (1-\gamma/6)^{-2} 
            \int_0^\infty dr_1 r_1^2 D^{-3+\gamma/2}_{L(r_1/d_*)^2} 
            \int_0^\infty dr_2 r_2^2 D^{-3+\gamma/2}_{L(r_2/d_*)^2} 
            r_{12}^{-\gamma}             
\eqno (4.5) 
$$ 
Performing here again the changes of variables (3.4), we obtain 
$$ 
w(\theta) = n_\Omega^{-2}\, \zeta_* (1-\gamma/6)^{-2} 
            d_*^{6-\gamma} \theta^{1-\gamma} 
            \int_0^\infty dq\, q^4 D_{L_*q^2}^{-6+\gamma} 
            \int_{-\infty}^{+\infty} du\, (q^2+u^2)^{-\gamma/2} 
\eqno (4.6) 
$$ 
and, using eq.~(2.7), we return to 
$$ 
w(\theta) = {\tilde A}_\gamma \left( {\tilde r}_0 \over d_* \right)^\gamma  
            \theta^{1-\gamma} 
\eqno (4.7) 
$$ 
provided that 
$$ 
{\tilde r}_0^\gamma {\tilde A}_\gamma  
= c_\gamma { \int_0^\infty dq\, q^{5-\gamma}  D^{-6+\gamma}_{L_*q^2}  
             \over [ \int_0^\infty dq\, q^2 D^{-3}_{L_*q^2} ]^2 } 
= 2\, c_\gamma \, L_*^{\gamma/2}  
           { \int_0^\infty dL\, L^{2-\gamma/2} D^{-6+\gamma}_L   
             \over [ \int_0^\infty dL\, \sqrt{L} D^{-3}_L ]^2 } 
\eqno (4.8) 
$$ 
($c_\gamma$ is defined in eq.~3.7). 
 
Besides of finding the same scaling properties as in the case of 
$L$--independent $r_0$, we can now try to answer a practical question: If we 
interpret angular results, possibly because of the scaling $w \propto  
d_*^\gamma$, as originating from a $L$--independent $r_0$ and work out the  
$r_0$ value, what value is obtained? 
 
This question is answered by requiring that 
$$ 
{r}_0^\gamma {A}_\gamma = {\tilde r}_0^\gamma {\tilde A}_\gamma  
\eqno (4.9) 
$$ 
and, using eqs.~(3.6) and (4.8), this yields 
$$ 
r_0^\gamma = { \int_0^\infty dL\, L^{2-\gamma/2}\, D_L^{-6+\gamma} 
            \over \int_0^\infty dL\, L^{2-\gamma/2}\, D_L^{-6} } ~, 
\eqno (4.10) 
$$ 
which is the {\sl apparent} value of $r_0$. 
 
Eq.~(4.10) can be used either with simulations, working out $D_L$ 
from them and performing numerically the two integrations, or with analytical
expressions. In the case of the Schechter law (2.1), using the function
$U(\lambda)$ defined in eq.~(2.3), it is easy to see that 
$$ 
r_0 = \phi_*^{1/3} \left[  
 \int_0^\infty d\lambda\, \lambda^{2-\gamma/2} U^{2-\gamma/3}_\alpha (\lambda) 
 \over \int_0^\infty d\lambda\, \lambda^{2-\gamma/2} U^2_\alpha (\lambda) 
 \right]^{1/\gamma}  ~.
\eqno (4.11) 
$$
The expression in square brackets can be turned into:
$$
[\Gamma(\alpha)]^{\gamma/3} v(\alpha,\gamma)
{ \int_0^\infty dx\, \exp(-x)\, x^{e_1} \left(\int_0^\infty dt\, \exp(-t)\,
{t^{1-\alpha} \over x/(2-\gamma/3) + t}\right)^{2-\gamma/3}
\over 
\int_0^\infty dx\, \exp(-x)\, x^{e_2} \left(\int_0^\infty dt\, \exp(-t)\,
{t^{1-\alpha} \over x/2 + t} \right)^{2}}
$$
with $e_1 = 2-\gamma/2+(2-\gamma/3)(1-\alpha)$ and
$e_2 = 2-\gamma/2+2(1-\alpha)$, while $v(\alpha,\gamma)
= 2^{e_2+1}/(2-\gamma/3)^{e_1+1}$.
Using a Gauss-Laguerre integration algorithm, double integrals can be then
turned into double summations and analytically evaluated. For the
narrow $\alpha$ interval, for which such expression is needed, however,
we can verify that the linear expression
$$ 
r_0/\phi_*^{1/3} = c_1*\alpha+c_2
\eqno (4.12)
$$
is better approximated than 0.2$\, \%$.

\begin{table}
\begin{center}
\begin{tabular}{|l|l|l|}
{$\gamma$}       &    { $c_1$}    &  { $c_2$}  \\
     1.6         &     -1.95      &   3.29     \\
     1.8         &     -1.93      &   3.25     \\
     2.0         &     -1.91      &   3.20     \\
     2.2         &     -1.88      &   3.15     \\
\end{tabular}
\label{tab1}
\end{center}
\end{table}

In Table 1 we give $c_1$ and $c_2$ for the relevant $\gamma$ interval.
Using such values it is easy to see that the discrepancy between 
the correlation length $r_0$
and the average distance between nearby objects is expected to be
$\mincir 10\, \%$. Such result, clearly, depends on the use of the
Shechter law (2.1), which also allows a confortable numerical integration.
We have considered a few other expressions, e.g. the expression
deduced from the Press \& Schecter (PS) cluster mass function, assuming
a constant spectral index $n_t$ for the transfered fluctuation spectrum 
and $L \propto M^2$. In most such cases, numerical integration only is
possible and integration details are cumbersome. For the PS case, however,
the expression (4.12) give results approximated up to some percents,
provided that we assume $\phi_* \equiv 2 \pi^{-1/2} n_t \nu^*$, where
$\nu^* = \rho_m/M_*$ ($\rho_m$ is the average matter density and $M_*$
is the mass attributed to the cluster with typical luminosity $L_*$).

\section{Final remarks}

According to Peebles (1980a), the scaling relation (3.5) played an important
role in testing that the angular correlations of galaxies in the catalogs 
do reflect the presence of a uniform spatial galaxy clustering, rather
than something else, e.g. systematic errors due to patchy obscuration
in the Milky Way. This work does not contradict such statement, corresponding
to the point (ii) at the beginning of section 2. However, Hauser and
Peebles (1973), in their seminal work on cluster clustering, made use of
the scaling relation (3.5), as a test for the cluster correlation length
$r_c \sim 30\, h^{-1}$Mpc they obtained for Abell clusters. This test,
as we showed, has only a partial validity.

For the sake of curiosity, let us outline that, if the galaxy luminosity
function values, $\phi^* \simeq 10^{-2}$ and $\alpha \simeq 1.2$
are used in eq.~(4.11), we obtain a galaxy correlation length
$r_g \simeq 6\, h^{-1}$Mpc, for $\gamma$ values in the interal 1.6--1.9.
This output should not be overstressed, as it has been known for a long
time that the average distance between nearby galaxies $D_{g}$ and the
galaxy correlation length $r_g$ have close values. In the case of galaxies,
there is little evidence of a linear relation between $r_g$ values
and $D_g$ values for intrinsical magnitude limited samples.
However, as galaxy luminosities essentially arise from stellar populations,
they may be subject to significant variations, as their stellar and
dust contents evolve on time scales certainly longer than observational times
(see, e.g., Silva et al 1999). It is possible that the proximity of $r_g$
and $D_g$ indicates that, averaging luminosities over suitable time scales,
a relation similar to clusters holds also for galaxies.

In the case of galaxy clusters, Dalton et al (1997) investigated the scaling
properties of the angular functions for two different samples with
apparent magnitudes $m > 18.94$ and $19.3 < m < 19.5$, for which they had
obtained suitable $U(\lambda)$ expressions. Assuming a constant spatial
clustering length $r_c$, they obtained that the predicted scaling properties
hold, while $\gamma = 2.14$ and $r_c = 14.3\, h^{-1}$Mpc. According to 
this work, finding a fair scaling does not require a constant $r_c$ and, 
henceforth, any conclusion based on its existence is somehow premature.

\end{document}